\documentclass[aps,pra,twocolumn,superscriptaddress,showpacs,floatfix]{revtex4-1}

\usepackage{amssymb,amsfonts,amsmath}
\usepackage{units} 
\usepackage{relsize} 

\usepackage{epsfig}
\usepackage{graphicx}

\makeatletter
\DeclareRobustCommand{\genericinterval}[2]{%
  \@ifstar{\genericinterval@star{#1}{#2}}{\genericinterval@nostar{#1}{#2}}}
\newcommand{\genericinterval@star}[4]{\mathopen{}\mathclose{\left#1#3,#4\right#2}}
\newcommand{\genericinterval@nostar}[4]{\mathopen{#1}#3,#4\mathclose{#2}}

\makeatother

\begin{document}
\newcommand{\vq}{\mathbf{q}}
\newcommand{\vp}{\mathbf{p}}
\newcommand{\vP}{\mathbf{P}}
\newcommand{\Wcm}{~Wcm$^{-2}\,$}
\newcommand{\Ea}{E_\mathrm{a}}
\newcommand{\Up}{U_\mathrm{p}}
\newcommand{\Upm}{U_{\mathrm{p},\max}}
\newcommand{\tp}{\tau_\mathrm{t}}
\newcommand{\tr}{\tau_\mathrm{r}}
\newcommand{\Tp}{T_\mathrm{p}}
\newcommand{\tg}{\tau_\mathrm{g}}
\newcommand{\vE}{\mathbf{E}}
\newcommand{\vA}{\mathbf{A}}
\newcommand{\ver}{\mathbf{r}}
\newcommand{\valpha}{\mbox{\boldmath{$\alpha$}}}
\newcommand{\vpi}{\mbox{\boldmath{$\pi$}}}
\newcommand{\Rep}{\mathrm{Re}\,}
\newcommand{\Imp}{\mathrm{Im}\,}
\newcommand{\AL}{A_\mathrm{L}}
\newcommand{\ve}{\hat{\mathbf{e}}}
\newcommand{\Ep}{E_\mathbf{p}}

\newcommand{\etal}{\emph{et al.~}}

\newcommand{\red}[1]{\textcolor{red} {\bf {#1}}}

\title{Streaking strong-field double ionization}

\author{M. K\"ubel}
\email{matthias.kuebel@uni-jena.de}
\affiliation{Joint Attosecond Laboratory, National Research Council and University of Ottawa, Ottawa, Ontario, Canada}
\affiliation{Department of Physics, Ludwig-Maximilians-Universit\"at Munich, D-85748 Garching, Germany}
\affiliation{Institute for Optics and Quantum Electronics, Friedrich-Schiller-Universit\"at, Jena, Germany}

\author{G. P. Katsoulis}
\affiliation{Department of Physics and Astronomy, University College London, London, United Kingdom} 

\author{Z. Dube}

\author{A. Yu.~Naumov}

\author{D. M. Villeneuve}

\author{P. B. Corkum}

\author{A. Staudte}
\affiliation{Joint Attosecond Laboratory, National Research Council and University of Ottawa, Ottawa, Ontario, Canada}

\author{A. Emmanouilidou}
\affiliation{Department of Physics and Astronomy, University College London, London, United Kingdom}

\begin{abstract}
Double ionization in intense laser fields can comprise electron correlations, which manifest in the non-independent emission of two electrons from an atom or molecule. However, experimental methods that directly access the electron emission times have been scarce. Here, we explore the application of an all-optical streaking technique to strong-field double ionization both theoretically and experimentally. We show that both sequential and non-sequential double ionization processes lead to streaking delays that are distinct from each other and single ionization. Moreover, coincidence detection of ions and electrons provides access to the emission time difference, which is encoded in the two-electron momentum distributions. The experimental data agree very well with simulations of sequential double ionization. We further test and discuss the application of this method to non-sequential double ionization, which is strongly affected by the presence of the streaking field.
\end{abstract}

\date{\today}

\maketitle

\section{Introduction}

	Photoionization is usually well described by assuming a single active electron \cite{Einstein1905}. The classic counter example in strong-field laser physics is Non-Sequential Double ionization (NSDI), see \cite{Faria2011,Becker2012,Bergues2015} for recent reviews. The term NSDI highlights the contrast to sequential double ionization (SDI), where the emission of two electrons from an atom or molecule can be understood as a sequence of two independent single-electron processes. In NSDI, the double ionization probability at moderate laser intensity is enhanced by several orders-of-magnitude with respect to SDI, which leads to the famous ``knee-shape" in the intensity-dependent yield curves for doubly charged ions of atoms and molecules \cite{LHuillier1983, Walker1994}. It has been suggested early on that electron correlations constitute the underlying reason for the double ionization enhancement \cite{Lambropoulos1985,LHuillier1986, Kuchiev1987, Fittinghoff1992}. Interestingly, deviations from the single active electron picture have also been reported for sequential double ionization \cite{Eichmann2000,Pfeiffer2011,Pfeiffer2011_NJP,Zhou2012}

The quest to understand NSDI has benefited from several groundbreaking experiments conducted approximately 20 years ago \cite{Moshammer2000,Weber2000a}. Highly differential measurements \cite{Moshammer2000,Weckenbrock2004,Staudte2007,Rudenko2007} have established that NSDI is accounted for by the laser-driven inelastic recollision of a first liberated electron \cite{Corkum1993}. Subsequent experimental and theoretical studies found that various mechanisms exist within this recollision picture \cite{Feuerstein2001,Liu2008,Camus2012}. For example, in the direct pathway, the recollision of the first electron directly promotes a bound electron into the continuum by an (e,2e)-type collision \cite{Weber2000a}. In the delayed pathway, on the other hand, the parent ion can be recollisionally excited and subsequently ionized (RESI) by the laser field \cite{Kopold2000,Feuerstein2001, Bergues2012}. Recently, it was shown that for small intensities slingshot-NSDI prevails the delayed pathway \cite{Katsoulis2018}. Other mechanisms invoke doubly excited states \cite{Prauzner-Bechcicki2005,Camus2012}. The various mechanisms can be characterized by the electron emission times, as has been done in many theoretical studies, e.g.~\cite{Ye2008,Emmanouilidou2009,Huang2013}. 
In experimental work, however, the emission times have not been directly accessed, even though their measurement would allow for a more direct comparison of experiment and theory. 

The emission time difference is a quantity with high significance in SDI, as well. Using the attoclock technique \cite{Eckle2008}, Pfeiffer et al.~\cite{Pfeiffer2011} measured the electron emission times within a laser pulse and found deviations from the single active electron approximation \cite{Pfeiffer2011,Pfeiffer2011_NJP,Zhou2012}. In the context of NSDI, however, the attoclock is not applicable, as the strongly elliptical polarization prevents recollision. Progress in accessing the emission time has been made by restricting NSDI to a single laser cycle by using few-cycle pulses with known carrier-envelope phase (CEP) \cite{Bergues2012,Camus2012,Bergues2015,Kubel2016_pra,Chen2017}.

In the present work, we explore the application of a streaking technique to the measurement of the electron emission times in sequential and non-sequential double ionization of argon. Figure 1 illustrates the experimental scheme. It uses an intense visible few-cycle pulse and a weaker, phase-stable mid-infrared (mid-IR) streaking field with perpendicular polarizations. The pulse duration of the visible pulse is shorter than a period of the mid-IR pulse, hence sampling its instantaneous vector potential \cite{Kubel2017_prl}. The visible pulse ionizes a target gas and accelerates the liberated electrons, potentially driving recollisions. In addition, photoelectrons are deflected by the mid-IR field, which does not ionize the target gas on its own. The mid-IR field imposes a momentum shift on the photoelectrons that depends on their time of ionization and the delay between the two pulses. Thus, measuring the momentum shift that the two electrons acquire from the acceleration in the mid-IR field provides information on their emission times. 

\begin{figure}[htp]
\centerline{\includegraphics[width=0.45\textwidth]{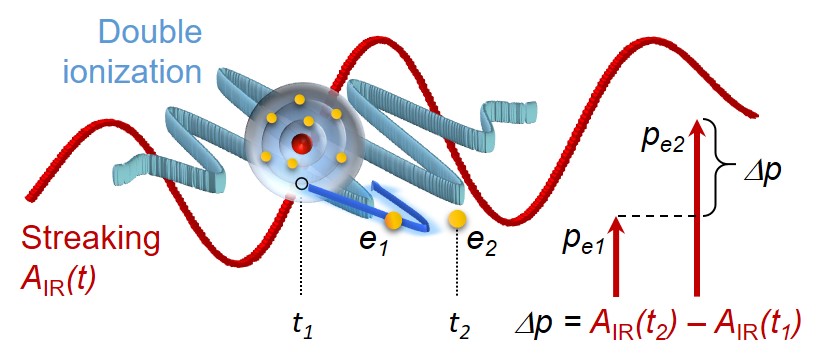}}
\caption{Schematic of the streaking experiment. An intense visible few-cycle pulse (cyan) drives double ionization of a rare gas atom in the presence of a mid-infrared streaking field (red). At times $t_1$ and $t_2$, electrons $e_1$ and $e_2$ are emitted, respectively. The laser-driven recollision of $e_1$ with the parent ion is indicated. The streaking field imposes different momentum shifts $A(t_1)$ and $A(t_2)$ on electrons $e_1$ and $e_2$, respectively. The momentum difference $\Delta p$ measures the electron emission time difference.}
\label{fig1}
\end{figure}

We present both experimental and theoretical results on streaking double ionization of argon. It is well known that both, direct and delayed ionization pathways, contribute to NSDI of argon \cite{Feuerstein2001} and sequential double ionization is achievable at relatively modest intensities \cite{Kubel2016_pra}. Thus, Ar represents an ideal testground to distinguish various double ionization mechanisms. We show that our streaking scheme allows two methods for accessing the electron emission time difference. 
The first approach measures the offset between the streaking curves obtained from the delay-dependent momentum distributions of singly and doubly charged ions. We show that the offset is sensitive to the underlying double ionization mechanism. In the second approach, coincidence detection of ions and electrons is utilized to measure both the electron momenta along the streaking field. As indicated in Fig.~\ref{fig1}, the momentum difference directly relates to the emission times of the electrons. 
We obtain excellent agreement between experiment and theory for SDI and discuss the application of the streaking scheme to NSDI.

\section{Methods}

\subsection{Experimental set-up}
The experiment relies on the STIER (subcycle tracing of ionization enabled by infrared) technique \cite{Kubel2017_prl}. The amplifier output from a 10\,kHz titanium:sapphire laser is split in two parts. The major part ($\approx 85\%$) is used to pump an optical parametric amplifier to obtain CEP-stable mid-infrared (mid-IR) idler pulses with a carrier wavelength of 2300\,nm. The minor part is focussed into an argon-filled hollow-core fibre. The spectrally broadened beam is compressed using chirped multilayer mirrors to obtain 5\,fs ``visible" pulses centered around 730\,nm. The polarization of the visible beam is adjusted to s-polarization using a broadband half-wave plate. Subsequently, it is recombined with the p-polarized mid-IR beam on a silicon mirror at 60$^\circ$ angle of incidence.

The recombined pulses are focussed into the center of a Cold Target Recoil Ion Momentum Spectrometer (COLTRIMS) where they intersect a cold ($T \approx 10$\,K) argon gas jet. In the focus, the intensity of the visible beam exceeds $10^{14}\,\mathrm{Wcm^{-2}}$, which suffices to doubly ionize neutral Ar. The intensity of the mid-IR pulses is of the order of $10^{13}\,\mathrm{Wcm^{-2}}$, low enough to avoid notable ionization of neutral argon by the mid-IR pulse on its own. Ions and electrons produced in the laser focus are guided onto time and position sensitive detectors using electric and magnetic fields. Their three-dimensional momenta are measured in coincidence and correlated with the delay between the visible and mid-IR pulses. The time-of-flight axis of the COLTRIMS, which provides the best momentum resolution for ions, is aligned with the polarization of the mid-IR pulse. While this allows for the measurement of the two-electron momentum distributions of double ionization along the mid-IR polarization, the ion momentum resolution in the perpendicular directions is not sufficient to accurately resolve the recoil momentum of Ar$^+$ and Ar$^{2+}$ ions. 

\subsection{Theoretical approaches}
\subsubsection{Non-Sequential Double Ionization}

For the NSDI calculations, a three-dimensional (3D) semi-classical model is employed, with the only approximation being in the initial state \cite{Emmanouilidou2008}. 

The laser field is described by
\begin{align}
\vec{E}(t) =& E_{\mathrm{VIS}}  f(t,\tau_\mathrm{VIS}) \cos\left( \omega_\mathrm{VIS}t +\phi \right) \hat{z} \nonumber \\
+&  E_\mathrm{IR} f(t+\Delta t,\tau_\mathrm{IR}) \cos\left( \omega_\mathrm{IR}({t+\Delta t})  \right) \hat{x}, 
\label{eq:laser field}
\end{align} 

with the envelope 
\begin{equation}
\mathrm{f(t,\tau)} = \exp\left(-2\log2 \left(\mathrm{\frac{t}{\tau}}\right)^2 \right).
\end{equation}

For the first, visible pulse, we use $\tau_\mathrm{VIS} = 5\,\mathrm{fs}$, $\omega_\mathrm{VIS} =  0.061\,\mathrm{a.u.}$ (i.e. $\lambda_\mathrm{VIS} = 750\,\mathrm{nm}$), $E_\mathrm{VIS} = 0.09\,\mathrm{a.u.}$ $(I = 3 \times 10^{14}\mathrm{Wcm^{-2}})$. For the second, mid-IR pulse, $\tau_\mathrm{IR} =  40\,\mathrm{fs}$, $\omega_\mathrm{IR} = \omega_\mathrm{VIS} / 3 $ ($\lambda_\mathrm{IR} = 2250\,\mathrm{nm}$), and $E_\mathrm{IR} =  0.03\,\mathrm{a.u.}$ $(I = 3 \times 10^{13}\mathrm{Wcm^{-2}})$ is used. The carrier-envelope phase (CEP) of the visible pulse is denoted by $\mathrm{\phi}$. The time delay between the two fields is given by $\mathrm{\Delta t}.$ Atomic units are used throughout this work unless otherwise indicated. 

One electron (electron 1) exits the field-lowered Coulomb barrier along the direction of the total laser field at time $t_0$ and is computed using parabolic coordinates \cite{Hu1997}. The ionization rate is obtained using the Ammosov-Delone-Krainov (ADK) formula \cite{Landau,Delone1991}. To identify the initial ionization time $t_0$, we employ importance sampling using the ionization rate as the distribution \cite{Rubinstein2016}. We do so in the time interval $[-\mathrm{\tau_{VIS}},\mathrm{\tau_{VIS}}]$. We have verified that the ionization rate drops to zero outside this time interval.


Upon tunnel ionization, we set the momentum of electron 1 along the direction of the laser field equal to zero, while the transverse momentum is given by a Gaussian distribution \cite{Landau,Delone1991}. This assumption was verified experimentally for strongly driven Ar \cite{Fechner2014}. The rest of the formulation, that is, the initial state of the initially bound electron (electron 2) and the time propagation is classical. The former is described by employing a microcanonical distribution \cite{Abrines1966}. The latter is achieved by solving Hamilton's equations of motion for the three-body system with the nucleus kept fixed. We fully account for the Coulomb singularities by using regularized coordinates \cite{Kustaanheimo1965,Heggie1974}.

Our results are obtained by averaging over twelve CEP values from $\phi = 0^\circ$ to $\phi = 330^\circ$ in steps of $30^\circ,$ for the visible field. The time delay between the pulses is varied with a step size of $0.1\, T_\mathrm{IR} = 2 \pi / \omega_\mathrm{IR}$ in the time interval $[-8,8]\, T_\mathrm{IR}.$ After propagating for a \emph{long time} we register the double ionization (DI) events. Next, we label the mechanism for each DI event according to the time difference from the recollision time $t_\mathrm{rec}$ to the emission time $t_{1/2}$ of electron 1 and electron 2, respectively. For the direct pathway the re-colliding electron 1 transfers enough energy to ionize electron 2 and both electrons escape shortly after re-collision. For the delayed pathway, electron 1 transfers enough energy so that only one electron (either electron 1 or electron 2) is emitted after re-collision and the other one transitions to an excited state and ionizes at a later time with the assistance of the laser field \cite{Kopold2000,Feuerstein2001} or with the assistance of the the nucleus and the field \cite{Katsoulis2018}. In practice, we register a DI event as direct or delayed depending on whether the time difference $ | t_\mathrm{rec}-t_i |$  is less than a small time interval $t_\mathrm{diff}$ for both electrons or for only one, respectively \cite{Chen2017}. We select $t_\mathrm{diff} = 1/10 T_{\omega_\mathrm{VIS}}. $

\subsubsection{Sequential Double Ionization}
For the calculations of single ionization and SDI we use a two-dimensional classical trajectory model \cite{Kubel2016_pra,Kubel2017_prl}, which allows for two tunnel ionization steps. The laser field is described by equation \ref{eq:laser field}. The ionization rates of Ar and Ar$^+$ are calculated using the rates proposed in Ref.~\cite{Tong2005}. The electron yields per time step are calculated by solving appropriate rate equations \cite{Kubel2016_pra}.

Liberated electrons are initialized at the tunnel exit along the instantaneous electric field vector. We set the initial momentum along $\hat{z}$ to zero and use a small, random momentum along $\hat{x}$. This approximation is reasonable, as the visible field is substantially stronger than the mid-IR field. The electrons are then propagated under the influence of the total laser field and a softcore coulomb potential $V_c = 1/(r+\alpha)$, where $r = \sqrt{x^2+z^2}$ and $\alpha = 1\,\mathrm{a.u.}$. 
The two electron spectra are generated for trajectories with positive final energy, weighed by the total ionization probability. The calculations are repeated for eight values of $\phi$ where for each value of $\phi$, $\Delta t$ is varied with a step size of 0.2\,fs.

\section{Results and Discussion}

\subsection{Effect of the streaking field on double ionization}

\begin{figure}[htp]
\centerline{\includegraphics[width=0.5\textwidth]{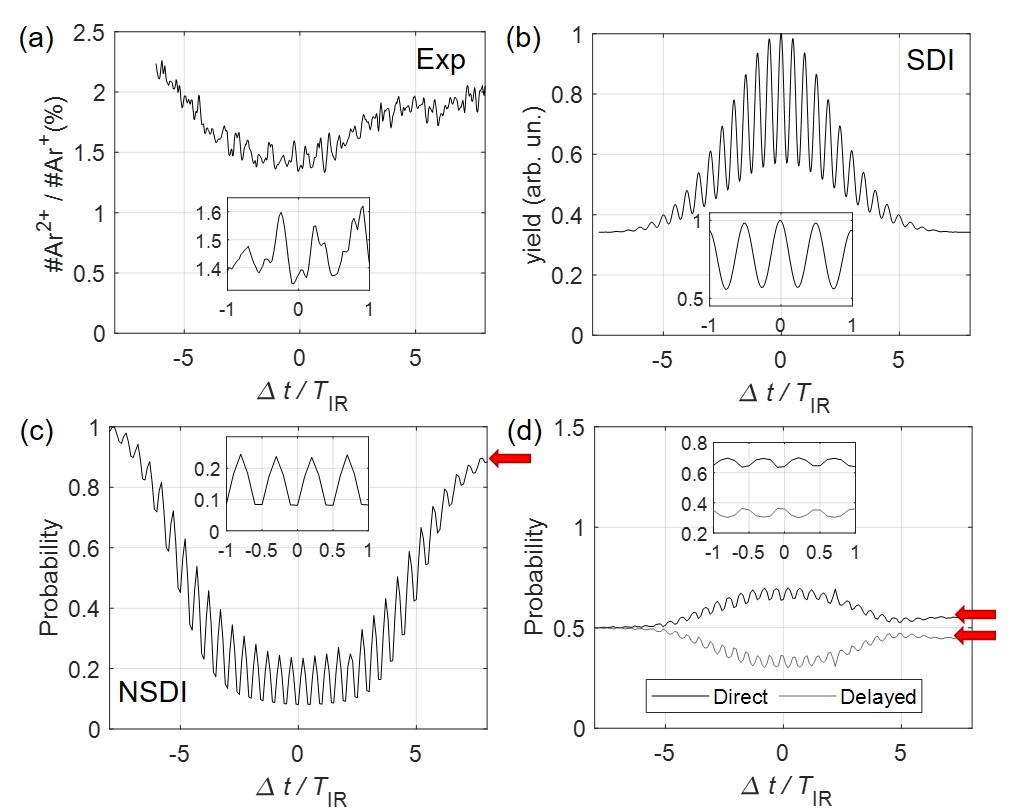}}
\caption{Delay dependence of the double ionization probability. The experimental data (a) shows the measured ratio of the Ar$^{2+}$ and Ar$^+$ yields. The inset enlarges the sub-cycle oscillations of the double ionization probability. Numerical results are presented for (b) SDI and (c) NSDI pathways. At $\Delta t = 0$, enhancement is obtained for SDI, and suppression is obtained for NSDI. Panel (d) shows the delay dependence of the share of direct and delayed NSDI mechanisms. The red arrows in panels (c) and (d) indicate the respective values in the absence of the mid-IR field. 
}
\label{fig2}
\end{figure}

Figure \ref{fig2}(a) shows the delay dependence of the measured ratio of Ar$^{2+}$ and Ar$^+$ yields. In the temporal overlap of the pulses, a substantial suppression of double ionization is observed. Notably, the double ionization probability oscillates with the half period of the mid-IR field. The lower value of the measured Ar$^{2+}$ yield at late delays compared to early delays is attributed to a low-intensity tail of the mid-IR pulse (see below), which slightly suppresses double ionization. 

The calculated delay-dependent SDI yield is presented in Figure \ref{fig2}(b). Contrary to the experiment, double ionization is enhanced in the overlap of visible and mid-IR pulses. In particular, a yield maximum is obtained at zero delay, where the field maxima of the mid-IR pulse overlaps with the center of the visible pulse. This suggests that the enhancement is due to the increased intensity in the pulse overlap. 

Figure \ref{fig2}(c) shows that the NSDI yield is strongly suppressed by the mid-IR field and exhibits a minimum at zero delay. This suppression can be understood by considering the effect of the orthogonally polarized mid-IR field on the recollision along the visible pulse polarization, which is necessary for NSDI. It is analogous to the well-known suppression of recollision in elliptically polarized laser fields. 
The fast oscillations of the double ionization probability can be explained as follows. First, the interval between tunnel ionization and recollision in the visible field amounts to approximately 1.6\,fs, less than a quarter-cycle of the mid-IR field. Thus, if tunnel ionization takes place near a zero-crossing of the mid-IR field, the net effect of the mid-IR field on the recollision trajectory is small and recollision remains possible. However, if ionization occurs in the vicinity of a field maximum, the mid-IR field accelerates the electron perpendicularly to the visible polarization, thus inhibiting recollision. 

The suppression of the NSDI yield, seen in Fig.~\ref{fig2}(c), is much stronger than in the experimental data presented in Fig.~\ref{fig2}(a). The weak suppression in the experimental data can be explained by contributions of SDI to the Ar$^{2+}$ yield. This delivers an explanation for the small amplitude of the measured yield oscillations, as the yield oscillations of NSDI and SDI are out of phase (cf.~the insets of Figs.\ref{fig2}(b,c)). The interpretation that both SDI and NSDI contribute to double ionization in the experiment is supported by additional experiments, where we have observed weaker suppression of double ionization at higher intensity and stronger suppression at lower intensity.

Interestingly, our NSDI simulations show a larger double ionization yield when the visible pulse precedes the mid-IR pulse, $\Delta t \approx -8\,T_\mathrm{IR}$ than when it succeeds the mid-IR pulse, $\Delta t \approx +8\, T_\mathrm{IR}$. In the latter case, the double ionization yield assumes the same value as in the single-pulse case, indicated by the red arrow. This suggests that the interaction of the atom with the visible pulse can produce excited ions that are subsequently ionized in the mid-IR field for $\Delta t \approx -8\,T_\mathrm{IR}$.

In Fig.~ \ref{fig2}(d), the effect of the mid-IR field on the contributions of direct and delayed mechanisms to the total NSDI yield is plotted. It can be seen that the mid-IR field suppresses delayed mechanisms with respect to direct mechanisms. When the visible pulse precedes the mid-IR pulse ($\Delta t \approx -8\,T_\mathrm{IR}$), the delayed pathways are enhanced with respect to the direct pathway. This again agrees with the conjecture that the visible pulse can produce excited ions that are ionized by the mid-IR pulse at a later time.

\subsection{Streaking delays}

\begin{figure}[htp]
\centerline{\includegraphics[width=0.5\textwidth]{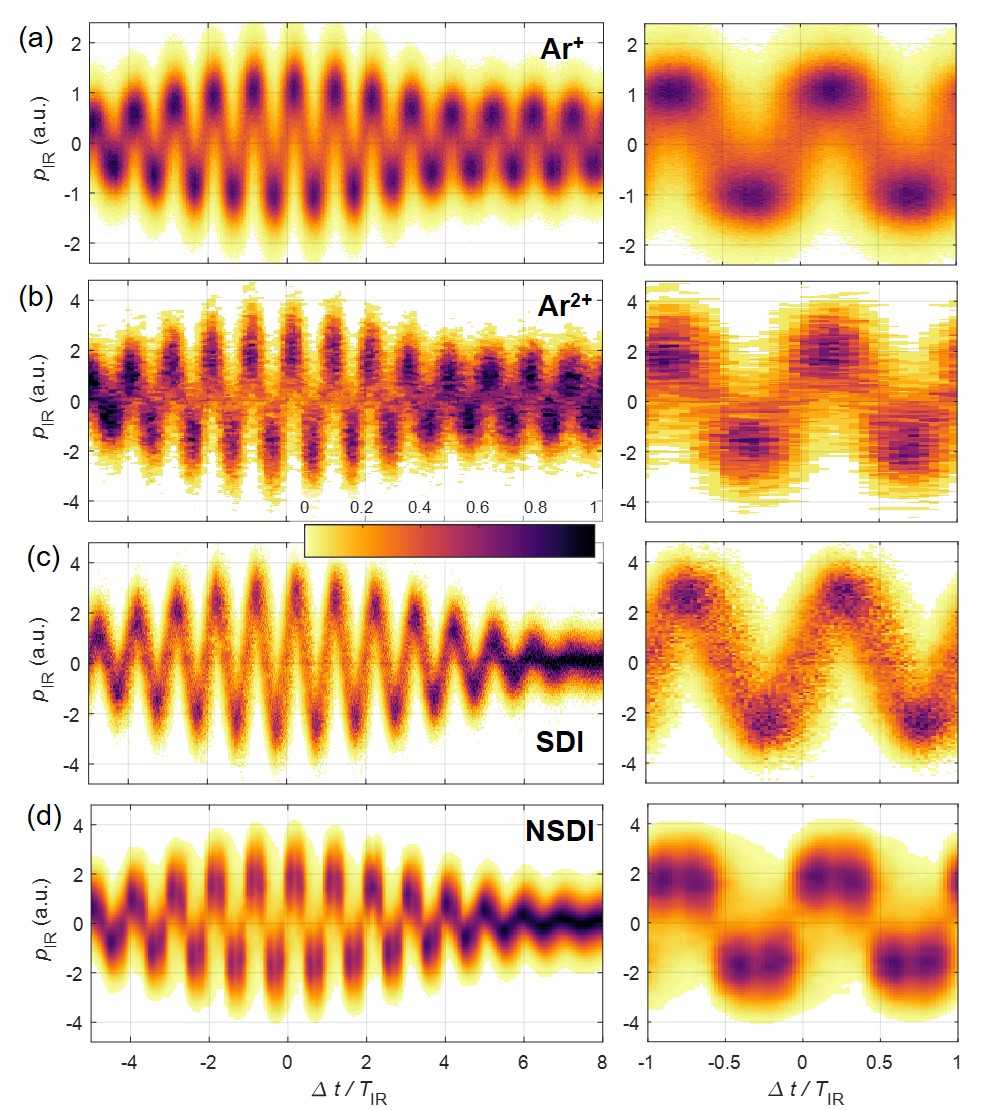}}
\caption{Delay-dependent momentum distribution measured for (a) Ar$^+$ and (b) Ar$^{2+}$ ions, and calculated for the Ar$^{2+}$ ion produced by (c) sequential double ionization (SDI) and (d) non-sequential double ionization (NSDI). To ease comparison of the traces, the yield in each delay bin is normalized to one. The small panels show zoom-ins around zero delay. The laser intensities for both experiment and theory are $I_\mathrm{VIS} = 3.0 \times 10^{14}\,\mathrm{Wcm^{-2}}$ and $I_\mathrm{IR} = 3.0 \times 10^{13}\,\mathrm{Wcm^{-2}}$.} 
\label{fig3}
\end{figure}

Figure \ref{fig3} shows experimental and numerical results for the streaking traces of Ar$^+$ and Ar$^{2+}$ ions. All streaking traces resemble the vector potential of the mid-IR pulse, which is a consequence of the short pulse duration of the visible pulse that confines ionization to a fraction of the mid-IR cycle. For delays $\Delta t > 4 T_\mathrm{IR}$, the amplitude of the momentum oscillation remains nearly constant over several cycles. This low-intensity tail of the mid-IR pulse is attributed to uncompensated spectral phase and is absent in the simulations, where Fourier-transform-limited pulses have been used. 

The streaking traces for Ar$^{2+}$ and the NSDI calculations exhibit pronounced yield maxima at the amplitudes of the momentum oscillations, and rather weak signal in between. This shape contrasts earlier measurements where the polarization vectors of visible and mid-IR fields were parallel \cite{Kubel2017_prl}. The maxima indicates that electron emission mostly occurs near the zeros of the streaking field. For NSDI, this is ascribed to the necessity of recollision, which is only possible around the zeros of the streaking field, as discussed above. 

The oscillations in the Ar$^{2+}$ momenta have twice the amplitude of those in the Ar$^+$ momenta. This is expected since the Ar$^{2+}$ momentum is the sum momentum of two electrons, whereas the Ar$^+$ momentum results from the emission of only one electron. It also indicates, however, that both electrons are mostly emitted within a time window that is small compared to the optical cycle of the mid-IR field ($T_\mathrm{IR} = 7.7\,\mathrm{fs}$).
Below, we analyze the measured and simulated streaking traces in detail by extracting the mean momentum at each time delay. 

\begin{figure}[htp]
\centerline{\includegraphics[width=0.5\textwidth]{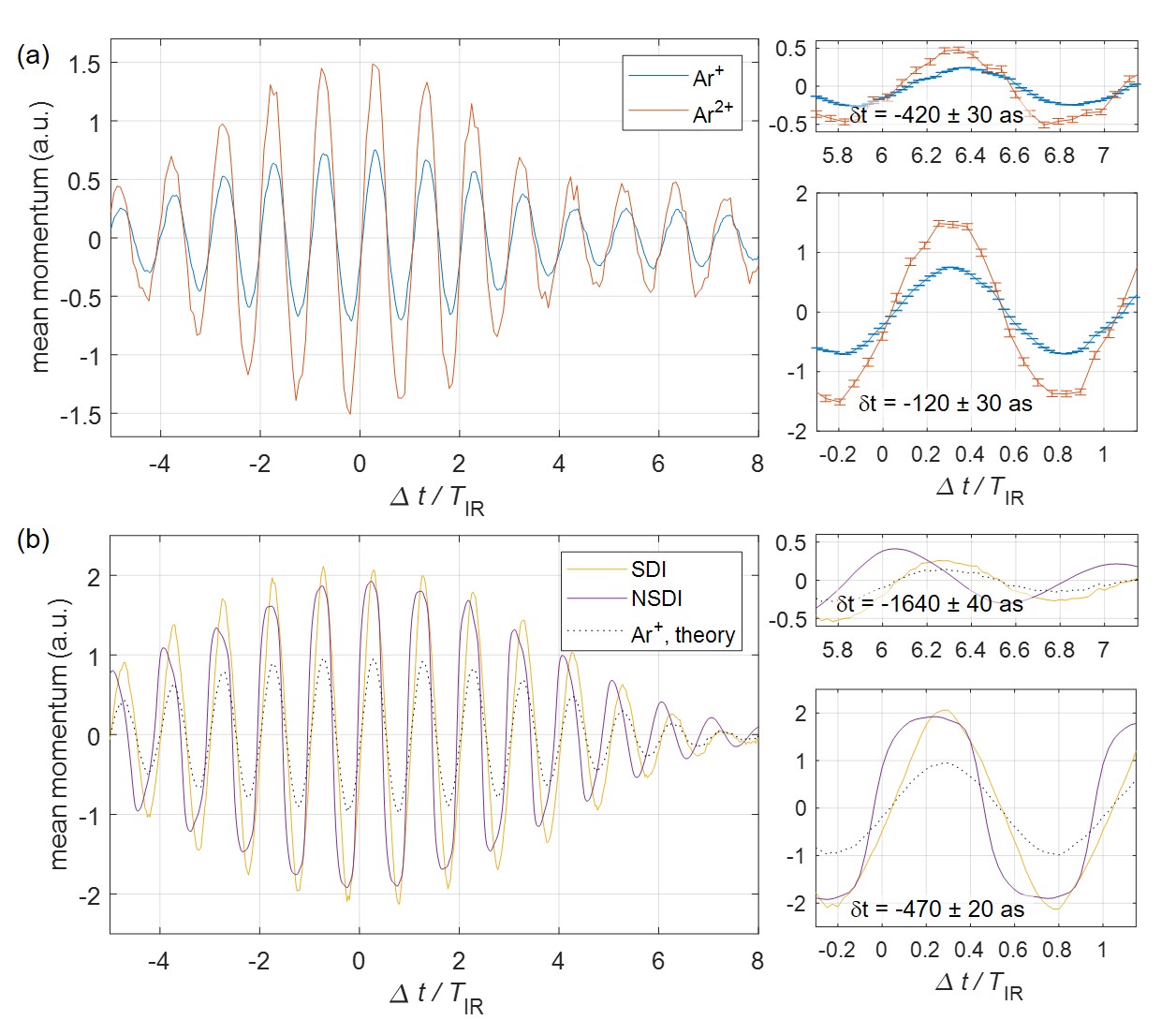}}
\caption{Streaking delays for double ionization mechanisms. Shown is the delay-dependent mean momentum of (a) the measured streaking traces for Ar$^+$ and Ar$^{2+}$ presented in Figs.~\ref{fig3}(a,b). In the right panels, the delays between the measured Ar$^+$ and Ar$^{2+}$ curves are extracted in the plotted regions. Errorbars are standard error of the mean. (b) Same as (a) for the calculated streaking traces for SDI and NSDI presented in Figs.~\ref{fig3}(c,d). Additionally, numerical results for single ionization of Ar are plotted. In the right panels, the delays between the NSDI curve and Ar$^+$ curves are extracted in the plotted regions. The offset between the SDI and Ar$^+$ curves is $50\pm 40\,\mathrm{as}$ at the center and $-80 \pm 30\,\mathrm{as}$ at the tail of the IR pulse.
}
\label{fig4}
\end{figure}

Figure \ref{fig4} shows the delay-dependent mean momenta extracted from the streaking traces shown in Fig.~\ref{fig3}. We investigate the two-electron emission dynamics by analyzing the delays between the curves for double and single ionization. 
For the experimental data presented in \ref{fig4}(a), we find that the delay offset between the Ar$^+$ and Ar$^{2+}$ curves varies throughout the mid-IR pulse. At the center of the mid-IR pulse, an offset of 120\,as is obtained that increases to 420\,as at the low-intensity tail of the mid-IR pulse. The dependence of the streaking delay on the mid-IR intensity is another signature of how the cross-polarized mid-IR field affects the double ionization dynamics.

Fig.~\ref{fig4}(b) shows the computational results for Ar$^+$ and Ar$^{2+}$ produced by either SDI or NSDI mechanisms. The Ar$^+$ and SDI curves exhibit only small offsets below 100\,as with respect to each other. For the offset between single ionization and NSDI, however, we observe a qualitatively similar behaviour as observed experimentally, i.e. the offset increases for decreasing intensity. This is consistent with the suppression of delayed NSDI at maximum overlap, as seen in Fig.~\ref{fig2}(d).
However, the computed delay shifts between Ar$^+$ and NSDI are approximately four times larger than the ones measured for the shift between the Ar$^+$ and Ar$^{2+}$ curves. This discrepancy further suggests significant contributions of SDI to the experimental signal. 


We have shown above that the shift between the streaking curves measured for single and double ionization is distinct for SDI and NSDI mechanisms. To discuss the relationship between these offsets and the emission time difference of the first and second electron, we approximate the delay-dependent electron momenta involved in double ionization by the separately measured delay-dependent momenta for single ionization (i.e.~Ar$^+$), and allow for a delay offset $\delta t_n$ ($n \in [1,2]$):
\begin{align}
\overline{p_{Ar^{2+}}}(\Delta t) &= -(\overline{p_1}(\Delta t) + \overline{p_2}(\Delta t)) \nonumber\\
 &\approx \overline{p_{Ar^+}}(\Delta t + \delta t_1) + \overline{p_{Ar^+}}(\Delta t + \delta t_2), 
\end{align} \label{eq:emission_time_difference}
where $\overline{p_1}$ and $\overline{p_2}$ are the delay-dependent momenta of the first and second electron, respectively. This formulation suggests that the streaking delay probes the mean emission time of the two electrons, but the emission time difference is not accessible.

The SDI simulations, where the first and second electron can be separated, justify this approximation. However, they also show that the delay offset for both electrons is significant: the first electron in SDI is typically emitted earlier, $\delta t_1 = 0.6\,\mathrm{fs}$, and the second electron later, $\delta t_2 = - 0.5\,\mathrm{fs}$, than the mean emission time for single ionization. These values result in the small net offset between single ionization and SDI, even though the average emission time difference between the first and second electron is on the order of 1\,fs. 

In NSDI, the first electron is emitted at the initial ionization time $t_0$, before recolliding with the parent ion, approximately 1.6\,fs (roughly two-thirds of a laser period) later. However, it has been shown in the context of NSDI, that the memory of the initial ionization is lost after recollision \cite{Camus2012}. Therefore, it is reasonable to interpret the final emission time of the first electron, $t_1$, as relevant for the streaking delay. The second electron is then emitted shortly after. The streaking delay shift of -1640\,as between NSDI simulations and single ionization obtained at late delays in Fig.~\ref{fig4}(b) agrees with this notion. Indeed, the simulations yield $\delta t_1 = 1.3\,\mathrm{fs}$, and $\delta t_2 = 1.9\,\mathrm{fs}$ at $\Delta t = 6T_2$.  At the center of the IR pulse, $\Delta t =0$, however, we obtain values of $\delta t_1 = 2.1\,\mathrm{fs}$, and $\delta t_2 = 3.0\,\mathrm{fs}$, which does not agree with the retrieved streaking delay of -470\,as. We attribute the discrepancy to the strong impact of the IR field on the NSDI dynamics, such that the electron emission is no longer confined to the center of the visible pulse. Instead, it occurs in a range of times that exceeds an optical period of the presently used mid-IR field, which impedes the measurement of unique and meaningful streaking delays.

We have shown above that the streaking delay shifts between single and double ionization are sensitive to the emission times of the two electrons. However, since the emission of either electron can exhibit a delay with respect to single ionization the retrieval of the emission time \emph{difference} from the streaking delay shifts is generally not possible. In the following, we analyze the difference of the electron momenta in order to directly access the emission time difference of the two emitted electrons, as illustrated in Fig.~\ref{fig1}.

\subsection{Two-electron momentum spectra}

\begin{figure}[htp]
\centerline{\includegraphics[width=0.5\textwidth]{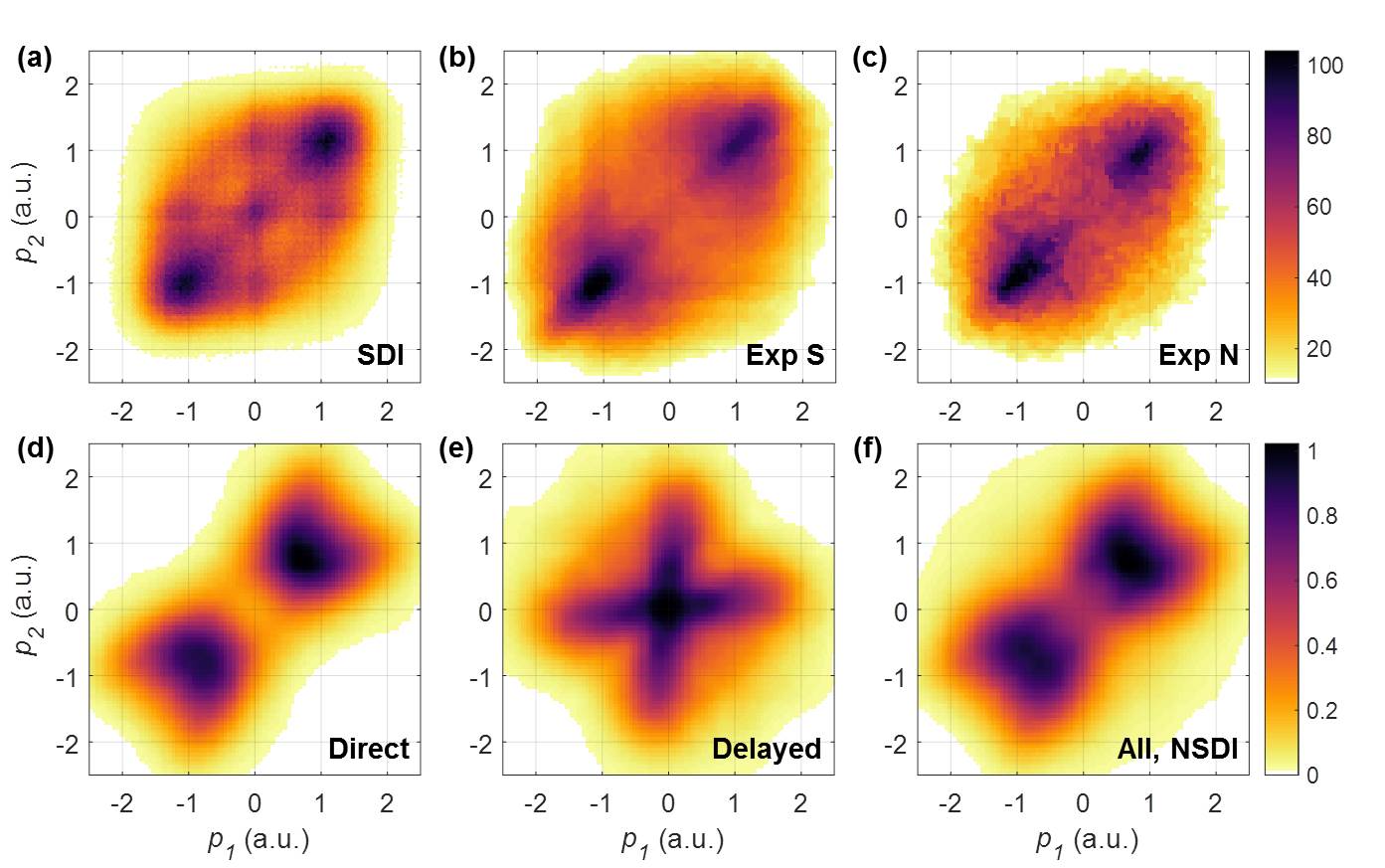}}
\caption{Two-electron coincidence spectra. The double ionization yield is plotted as a function of the momentum components along the mid-IR polarization of the two electrons. Numerical results for SDI in the delay range are shown in panel (a). Experimental results in the delay range are presented for (b) $I_\mathrm{VIS} = 3.5\times 10^{14}\mathrm{Wcm^{-2}}$ and (c) $I_\mathrm{VIS} = 2.0\times 10^{14}\mathrm{Wcm^{-2}}$. Numerical results for NSDI are presented in (d) for the direct pathway, (e) delayed pathways, and (f) all NSDI events. All plots are normalized to 1. The color bar at panel (c) applies to all experimental results, which were integrated in the delay range $[-2 T_\mathrm{IR},2 T_\mathrm{IR}]$. The color bar at panel (f) applies to numerical results, which were integrated in the delay range $[-T_\mathrm{IR},T_\mathrm{IR}]$.}
\label{fig5}
\end{figure}

In Fig.~\ref{fig5}, we present the two-electron momentum distributions for the momentum along the mid-IR polarization, obtained from two measurements at different intensities, as well as SDI and NSDI simulations. The distributions are symmetric with respect to the major diagonal, as the first and second electron are indistinguishable. Moreover, they are symmetric with respect to the minor diagonal, as the data has been integrated over a delay range corresponding to a few optical cycles, around the center of the mid-IR pulse. The main feature of the computational SDI results, shown in Fig.~\ref{fig5}(a), is a strong signal along the main diagonal with pronounced maxima in the first and third quadrants. It agrees well with the experimental results at two different intensities, that are presented in Figs.~\ref{fig5}(b) and (c). This suggests again that, at the present experimental conditions, double ionization in the overlap region is dominated by SDI.

The results of the NSDI calculations are shown in Figs.~\ref{fig5}(d-f). As seen, the direct and delayed pathways lead to distinct patterns in the electron momentum distributions along the streaking axis. Interestingly, both the ``bunny ears" \cite{Parker2006,Staudte2007, Rudenko2007, Emmanouilidou2008,Ye2008} for direct NSDI and the ``cross" \cite{Bergues2012} for delayed double ionization are reminiscent of results reported for single-pulse NSDI experiments. 

In contrast, the two-electron momentum distributions pattern obtained in single-pulse SDI experiments with linear polarization \cite{Weber2000a, Kubel2016_pra}, exhibit a featureless spot around the origin. 
In the present experiment, however, the mid-IR streaking enforces a strong correlation of the two electron momenta, that is seen in Figs.~\ref{fig5}(a-c), as it imposes nearly equal momentum shifts onto electrons emitted within the same half-cycle. For electrons produced in different half-cycles, larger momentum differences are obtained. As illustrated in Fig.tunnelling~\ref{fig1}, the momentum difference can provide access to the emission time difference.

\begin{figure}[htp]
\centerline{\includegraphics[width=0.5\textwidth]{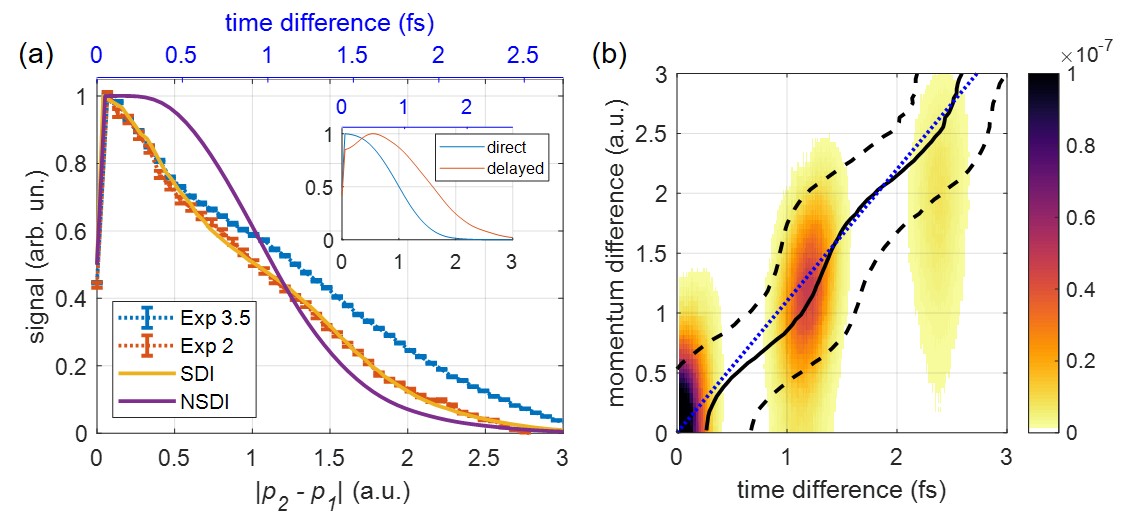}}
\caption{Momentum difference and emission time difference. (a) The momentum differences along the mid-IR polarization for the experimental and numerical results presented in Fig.~\ref{fig5}. For the experimental results, a homogeneous background of 10\% of the maximum signal has been subtracted. The inset shows the momentum difference distribution for direct (blue) and delayed (red) NSDI mechanisms. The time difference axis on top of panel (a) has been obtained from panel (b), where the calculated SDI yield is shown as a function of emission time difference and momentum difference of the two electrons. For clarity, only delay values where the visible pulse is in the vicinity of the zero crossing of the IR vector potential were used to obtain this plot. The solid black line represents the mean emission time for each momentum bin, the dashed lines indicate the standard deviation. The dotted blue line is a linear fit to the mean emission time, where the offset momentum was forced to zero. A linear fit was used to obtain the time difference axis in panel (a). }
\label{fig6}
\end{figure}

In Fig.~\ref{fig6}(a), we present the distribution of the momentum differences for the two-electron momentum spectra of Fig.~5. The shapes of both experimental results agree well with the SDI results. The strongest contributions come from electron pairs with small difference momenta, corresponding to small emission time differences. In addition, a clear shoulder around $|p2-p1| \approx 1.3\,\mathrm{a.u.}$ is visible. This corresponds to a time difference just over 1~fs, in reasonable agreement with the half-period of the visible field. Thus, the shoulder arises from electron emission in subsequent half-cycles. A shoulder that corresponds to time differences of a full optical cycle is not clearly visible. 

Remarkably, quantitative agreement between SDI calculations and experimental results obtained at $I_\mathrm{VIS} = 2.0 \times 10^{14}\,\mathrm{Wcm^{-2}}$ is obtained when a homogeneous background of 10\% is subtracted from the data. At the higher intensity of $I_\mathrm{VIS} = 3.5 \times 10^{14}\,\mathrm{Wcm^{-2}}$, the contributions at larger difference momenta become more significant. This corresponds to increased contributions of trajectories with higher emission time difference. At increased intensity, such contributions are expected and are consistent with the attoclock results reported in Ref.~\cite{Pfeiffer2011}.
In that work, the average emission time difference was reported to be smaller than expected from theory \cite{Pfeiffer2011}. This effect is not observed in our data. 

The NSDI results show a much broader distribution at momentum differences up to 1\,a.u.~than the experimental results, again suggesting that the experimental double ionization yield in the overlap region is dominated by SDI, even at $2.0 \times 10^{14}\,\mathrm{Wcm^{-2}}$. The inset of Fig.~\ref{fig6}(a) shows the calculated electron momentum differences for the delayed and direct mechanisms. As expected, the delayed mechanism leads to larger differences in the momenta. Using the conversion from momentum difference to time difference, we obtain $0.5\,\mathrm{fs}$ for the direct and $0.9\,\mathrm{fs})$ for the delayed mechanisms. The 0.4\,fs difference between delayed and direct mechanism agrees with the results of Ref.~\cite{Bergues2012}.

Fig.~\ref{fig6}(b) shows the relationship between momentum difference and emission time difference for the SDI simulations. A linear fit to the solid line is used to create the time difference axis in Fig.~\ref{fig6}(a). The standard deviations (dashed lines) indicate that, under the present conditions, the uncertainty in the retrieval of emission times amounts to $\approx \pm 400\,\mathrm{as}$. The precision could be improved by analyzing the momentum difference along the direction of the visible polarization, which can provide an unambiguous time to momentum mapping for small time delays. 


\section{Conclusion and Outlook}

We have demonstrated streaking of strong-field double ionization by a phase-stable mid-infrared pulse. The measurement of the streaking delays between single and double ionization is sensitive to the emission times of the first and second electron. Direct access to the emission time difference is possible through the momentum difference of the two electrons along the streaking mid-IR laser field. While our experiment has successfully probed the two-electron emission in SDI, NSDI has been strongly suppressed by the cross-polarized mid-IR field. The application of even longer wavelength as streaking fields will aid measuring the emission time difference in NSDI, as the same deflection amplitude can be achieved at lower intensity, where NSDI is not affected as strongly as in the present study. 

On another frontier, the increased double ionization yield observed in the NSDI calculations when the visible pulse precedes the IR pulse (see Fig.~\ref{fig2}) suggests that our experimental scheme is useful to probe excitation dynamics in strong fields, such as frustrated tunnelling ionization \cite{Nubbemeyer2008}. At long wavelengths, these highly excited atoms may be ionized more efficiently than by the visible field that created them.

\begin{acknowledgments}
We thank D.~Crane, R.~Kroeker and B.~Avery for technical support. This project has received funding from the EU’s Horizon2020 research and innovation programme under the Marie Sklodowska-Curie Grant Agreement No. 657544. Financial support from the National Science and Engineering Research Council Discovery Grant No. 419092-2013-RGPIN is gratefully acknowledged. A.E.  acknowledges the EPSRC grant no. N031326 and the use of the computational resources of Legion at UCL.

\end{acknowledgments}

\bibliography{di_stier_bib}

\end{document}